\documentclass[prb,preprint]{revtex4}%
\usepackage{graphicx}
\usepackage{amsmath}
\usepackage{ulem}
\usepackage{color}

\newcommand{\be}{\begin{equation}}
\newcommand{\ee}{\end{equation}}
\newcommand{\bea}{\begin{eqnarray*}}
\newcommand{\eea}{\end{eqnarray*}}
\newcommand{\bean}{\begin{eqnarray}}
\newcommand{\eean}{\end{eqnarray}}

\begin{document}

\draft
\title
{\bf Heat rectification effect of serially coupled quantum dots}

\author{ Yen-Chun Tseng$^{1}$,  David M T Kuo$^{1,2}$, Yia-chung Chang$^{3}$, and Yan-Ting  Lin$^{1}$,}
\address{$^{1}$Department of Electrical Engineering and $^{2}$Department of Physics, National Central
University, Chungli, 320 Taiwan}


\address{$^{3}$Research Center for Applied Sciences, Academic Sinica,
Taipei, 11529 Taiwan}

\date{\today}

\begin{abstract}
The nonlinear thermoelectric properties of serially coupled quantum
dots (SCQDs) embedded in a nanowire connected to metallic electrodes
are theoretically studied in the Coulomb blockade regime. We
demonstrate that the electron heat current of SCQDs exhibit a
direction-dependent behavior (heat rectification) in an asymmetrical
structure in which the electron Coulomb interactions are
significant. The phonon thermal conductivity of the nanowire is also
calculated, which is used to estimate the phonon heat current.
Finally, we discuss how to reduce phonon heat current to allow
observation of electron heat rectification behavior in the SCQD
junction system in low temperature regime ($\approx 3~K$).
\end{abstract}

\maketitle
Transport properties of a single semiconductor quantum dot (QD)
or nanostructure have been experimentally and theoretically studied
in the Coulomb blockade regime for application as a single
electron transistor due to its functionality of charge filtering.$^{1-5}$
It has been proposed that the QD system can be used as a qubit for quantum computation.$^6$
For the realization of solid state quantum computer, many experimental
studies have been devoted to the tunneling current of serially coupled
double quantum dots (DQDs).$^{7}$ The serially coupled DQDs can act as a spin
filter when the the Pauli spin blockade condition is met.$^7$ Recent
experimental works have extended DQDs to triple quantum dots (TQDs)
for studying multi electron spin blockade process and leakage
current arising from phonon assisted tunneling.$^{8-10}$ However,
there has been little study on the energy transfer of such
nanostructure junctions including DQDs and TQDs. The understanding of energy transfer
and heat extraction of the nanostructure junction is also crucial in the implementation of
solid state quantum register, because the heat accumulation will degrade
the performance of quantum computation.

Apart from that, solid state coolers and power generators at
nanoscale may be important in the integration of quantum device
circuits.$^{11}$ Unlike electronic nanodevices, it is still a challenge
to realize nanoscale solid state thermoelectric devices.$^{11}$
Up to date,  several theoretical studies on the
thermoelectric properties of nanostructures have been reported.$^{12-17}$These
studies have focused on the thermoelectric properties of
QD junctions in the linear response regime rather than nonlinear
response. To the best of our knowledge, the nonlinear electron heat transport properties of
serially coupled QDs (SCQDs) including DQDs and TQDs have not been
reported. Here, we demonstrate that DQDs and TQDs with
asymmetrical structures can exhibit direction-dependent electron heat
current in the absence of phonon heat current. Such a rectification
effect of electron heat current may be masked by the phonon heat
current, since phonon carriers dominate the contribution of
heat current in nanowires with large cross-section. Thus, only for
SCQDs embedded in a thin nanowire, such
heat rectification effect may be realized.

Here, we consider nanoscale semiconductor QDs, in which the energy
level separations are much larger than their on-site Coulomb
interactions and thermal energies. Thus, only one energy level for
each quantum dot needs to be considered. The extended Hubbard model
and Anderson model are used to describe the SCQD system connected
to the metallic electrodes.$^{15}$ Using the Keldysh-Green's
function technique [15,18], the charge and heat currents of a DQD
[as shown in the inset of Fig. 1(a)] are calculated according to
\begin{eqnarray}
J&=&\frac{2e}{h}\int d\epsilon {\cal T}(\epsilon)
[f_L(\epsilon)-f_R(\epsilon)],\\ Q&=& \frac{2}{h}\int d\epsilon
{\cal T}(\epsilon)(\epsilon-E_F-e\Delta V)
[f_L(\epsilon)-f_R(\epsilon)],
\end{eqnarray}
where ${\cal T}(\epsilon)\equiv ({\cal T}_{12}(\epsilon) +{\cal
T}_{21}(\epsilon))/2$ is the transmission coefficient.$^{15,16}$
$f_{L(R)}(\epsilon)=1/[e^{(\epsilon-\mu_{L(R)})/k_BT_{L(R)}}+1]$
denotes the Fermi distribution function for the left (right)
electrode. $\mu_L$ and $\mu_R$ denote the chemical potentials of the
left and right leads, respectively, with their average denoted by
$E_F=(\mu_L+\mu_R)/2$. $\Delta V=(\mu_L-\mu_R)/e$ is the voltage
across the SCQD junction. $T_{L(R)}$ denotes the equilibrium
temperature of the left (right) electrode. $e$ and $h$ denote the
electron charge and Planck's constant, respectively. ${\cal
T}_{\ell,j}(\epsilon)$ denotes the transmission function, which can
be calculated by evaluating the on-site retarded Green's function
(GF) and lesser GF [15]. The indices ${\ell}$ and j denote the
${\ell}$th QD and the $j$th QD, respectively. Based on the equation
of motion method, we can obtain analytical expressions of all GFs in
the Coulomb blockade regime. Details are provided in Ref. 15. In the
weak interdot limit ($t_c/U_{\ell} \ll 1$, where $t_c$ and
$U_{\ell}$ denote the electron interdot hopping strength and on-site
Coulomb interaction, respectively) the transmission function can be
recast into a simple closed-form.$^{16-17}$

To study the direction-dependent heat current, we let $T_L = T_0 +
\Delta T/2$ and $T_R = T_0-\Delta T/2$, where $T_0 = (T_L + T_R)/2$
is the average of equilibrium temperatures of two side electrodes
and $\Delta T = T_L -T_R $ is the temperature difference across the
junction. Because the temperature gradient can induce a significant
electrochemical potential difference $\Delta V$, now denoted by
$\Delta V_{th}$ (the thermal voltage), it is important to take into
account the shift of energy level ($E_{\ell}$) in each dot according
to the expression $\epsilon_{\ell} = E_{\ell} + \eta_{\ell}e\Delta
V_{th}$, where $\eta_{\ell}$ denotes the fraction of voltage
difference shared  by QD $\ell$. The value of $\eta_{\ell}$ depends
on the location, shape and dielectric constant of the QD. For
simplicity, we assume that $\eta_{\ell}$ is determined solely by the
QD location and the voltage difference is uniformly distributed
among QDs. This is a valid approximation when the dielectric
constants of the QD and the surrounding material are similar, which
leads to a uniform electric filed in the junction system. Let
$d_{\ell}$ denotes the center position of QD $\ell$ with respect to
the mid point of the junction and the separation of two electrodes
is $D$, then the electrostatic potential energy due to the uniform
electric field seen by an electron in QD ${\ell}$ is simply $V({\bf
r}-d_{\ell}\hat z)=[d_{\ell}+(z-d_{\ell})](-e\Delta V_{th}/D)$ ($z$
is along the direction of transport). For weak field and symmetric
wave function in each QD, the energy correction due to the linear
$(z-d_{\ell})$ term vanishes up to first order. Thus, we have
$\eta_{\ell}=d_{\ell}/D$. For the DQD junction, we assume $d_1=-d_2$
and $\eta_1=-\eta_2$.

We have numerically solved Eqs. (1) and (2) for SCQD junctions. We
first determine $\Delta V_{th}$ by solving Eq. (1) with J = 0 (the
open circuit condition) for a given $\Delta T$, $T_0$ and an initial
guess of the average one-particle and two-particle occupancy
numbers, $N$ and $c$ for each QD. Those numbers are then updated
according to Eqs. (5) and (6) of Ref. 16 until self-consistency is
established. Once $\Delta V_{th}$ is solved, we then use Eq. (2) to
compute the heat current. We have adopted the following physical
parameters $U_{\ell}=300\Gamma_0$ and $T_0=26\Gamma_0$. If we
adopted an energy unit $\Gamma_0=10\mu eV$, then $U_{\ell}=3 meV$,
and $T_0 \approx 3 K$.


Figure 1(a) shows the heat current ($Q$) as a function of
temperature difference ($\Delta T$) for a DQD with various values of
$\eta$, where $\eta\equiv \eta_1=-\eta_2$. The QD energy levels are
$E_A=E_F+20\Gamma_0+\eta e\Delta V_{th}$ and
$E_B=E_F+10\Gamma_0-\eta e\Delta V_{th}$. The inter-dot Coulomb
energy is $U_{\ell,j}=20\Gamma_0$. A nonlinear behavior of heat
current is clearly seen in Fig. 1(a), where the heat current in the
forward temperature bias ($\Delta T > 0$) is much larger than that
in the reversed temperature bias ($\Delta T < 0$). This nonlinear
behavior in heat current leads to significant thermal rectification
effect for the DQD junction system, which depends strongly on
$\eta$. When $\eta=0$, the electrochemical potential $\Delta V_{th}$
yielded by temperature bias will not shift the QD energy levels.
Consequently, the thermal rectification effect disappears even
though we have $E_A \neq E_B$. Fig. 1(b) shows the corresponding
Seebeck coefficient. $S_0=\Delta V_{th}/\Delta T$ denotes the
Seebeck coefficient in the linear response ($\Delta T \rightarrow
0$) regime. A negative $S_0$ value implies that electrons of the hot
side electrode diffuse to the cold side electrode via the energy
levels $E_{\ell}$ above $E_F$.$^{15}$ On the other hand, holes of
the electrodes are the major diffusion carriers if we have a
positive $S_0$. Holes are defined as the empty states which are
below the Fermi energy of electrodes. Note that the relationship
between the electron heat current and Seebeck coefficient is highly
nonlinear, it is hard to verify the electron heat rectification via
the measurement of $S$, even though $S$ is not influenced by the
phonon heat current. In this study we shall consider configurations
that lead to large electron heat current, which is in contrast to
the Pauli spin blockade configuration considered in Refs. 7 and 15.

To further clarify the rectification behavior shown in Fig. 1(a), we
show the heat current ($Q$) of the DQD junction for various interdot
Coulomb interactions ($U_{\ell,j}$) in Fig. 2(a). Other physical
parameters are the same as those in Fig. 1(a) for $\eta=0.3$. With
decreasing $U_{\ell,j}$, the magnitude of $Q$ is seriously reduced
for large forward temperature bias, and at the same time, the
rectification behavior is seriously suppressed. Once all electron
Coulomb interactions are excluded, the rectification behavior is
very small (not shown here). This implies that the heat current
rectification is directly related to the electron Coulomb
interactions, especially the interdot Coulomb interactions. Note
that there are eight configurations for a electron with spin
$\sigma$ to diffuse from the hot electrode to the cold electrode
[see Eq. (3) of Ref. 16]. To reveal which channel dominates the heat
current $(Q)$, the curve labeled by $p_3$ (which is calculated by
including the $p_3$ configuration alone) is also plotted in Fig.
2(a) (triangles line). In the reversed temperature bias, it matches
very well with the solid line (including eight configurations),
whereas its magnitude is larger than that of solid line in the
forward temperature bias. This is mainly because the $p_1$ channel
allows electrons of the cold electrode to transport heat to hot
side. From the results of curve labeled $p_3$, the heat current is
determined by the joint density of states arising from two poles
$\epsilon_1=E_F+U_{12}+20\Gamma_0+\eta \Delta V_{th}$ and
$\epsilon_2=E_F+U_{12}+10\Gamma_0-\eta \Delta V_{th}$, which are
separated by $10\Gamma_0+2\eta \Delta V_{th}$ in the small $t_c$
limit. This explains why the heat current in the forward temperature
bias is much larger than that in the reversed temperature. Fig. 2(b)
shows the corresponding Seebeck coefficients ($S$) for various
values of $U_{\ell,j}$.  We see that $|S|$ diminishes as the
interdot Coulomb interaction decreases, since the tunneling energy
levels become closer to the Fermi energy of electrodes and the
contribution due to holes increases.

To design a good thermal rectifier, it is important to have large
heat current and high rectification efficiency. Fig. 3 shows the
heat current, $\Delta V_{th}$ and rectification efficiency
($\eta_Q$) for different interdot hopping strengths. Other physical
parameters are the same as those in Fig. 1(a) for $\eta=0.3$. To
elucidate the rectification efficiency of QD junction system, we
define $\eta_Q=(Q(\Delta T_{F})-|Q(\Delta T_R)|)/Q(\Delta T_{F})$,
where $\Delta T_{F(R)}$ denote the forward (reversed) temperature
bias. We find that the rectification efficiency is suppressed when
$t_c$ increases (accompanying an increasing heat current). On the
other hand, $\Delta V_{th}$ is insensitive to the variation of $t_c$
even in the nonlinear response regime. In the linear response
regime, $S_0$ is very insensitive to $t_c$ as long as $t_c/U_{\ell}
\ll 1$.$^{16}$

To examine the effect of coupling between QDs and electrodes on
thermal rectification efficiency, Fig. 4 shows the heat current,
$\Delta V_{th}$ and rectification efficiency for different tunneling
rate values,  while keeping $\Gamma_L+\Gamma_R=6\Gamma_0$ and
$t_c=3\Gamma_0$. Other physical parameters are the same as those in
Fig. 1(a) for $\eta=0.3$. Unlike parallel QD cases, where very
asymmetrical tunneling rates ($\Gamma_L/\Gamma_R \gg 1$~or
$\Gamma_R/\Gamma_L \gg 1$) are required to observe thermal
rectification effect,$^{13}$ the thermal rectification behavior of
SCQDs is not sensitive to the coupling condition between QDs and
electrodes. Based on the results of Figs. (1)-(4), the heat current
rectification of SCQD is very robust. Ref. 19 attempted to observe
thermal rectification experimentally in a single QD system with two
levels by using the measurement of $S$ in the linear response regime
($\Delta T \rightarrow 0$). Our calculated results imply that it is
difficult to verify the existence of electron thermal rectification
in either the linear or nonlinear response regime via the
measurement of $S$ due to unclear relationship between $Q$ and $S$.

So far, the results of Figs. 1-4  are related to the heat
rectification of DQDs. To build a large temperature bias (large
$\Delta T$) across the junction, it is essential to consider large
number of coupled QDs in the system in order to reduce thermal
current arising from  phonons.$^{11,17}$ For simplicity, we consider
the case of a TQD junction as shown in the inset of Fig. 5, similar
to the structure considered in Ref. 9. The expressions of Eqs. (1)
and (2) are still valid for the TQD junction, whereas we need to
consider a more complicated transmission coefficient.$^{17}$ The TQD
junction has 32 configurations for an electron with spin $\sigma$ to
tunneling between two metallic electrodes.$^{17}$ Figure 5 shows the
heat current ($Q$), and heat rectification efficiency for different
physical parameters in the TQD junction with
$E_A=E_F+30\Gamma_0+0.3e\Delta V_{th}$, $E_B=E_F+20\Gamma_0$, and
$E_C=E_F+10\Gamma_0-0.3e\Delta V_{th}$. The heat current has a
remarkable enhancement for $t_c=3\Gamma_0 $ and $\Gamma=5\Gamma_0$.
If we adopt $\Gamma_0=10\mu eV$, then $U_{\ell}=3~meV$,
$U_{\ell,j}=0.15~meV$, $t_c=0.03~meV$, $\Gamma=0.05~meV$, and
$T_0=0.26~meV=3.12~K$ in Fig.~5. Based on these physical parameters,
it is possible to fabricate a realistic TQD junction system to
realize the heat rectification behavior as predicted theoretically.
As an example, we have calculated the above physical parameters for
an $In_{0.6}Ga_{0.4}As/GaAs$ QD junction system within the effective
mass approximation.$^{20}$ For a disk-shaped QD with radius
$R_0=25~nm$ and height $L_{0}=30~nm$, the intradot Coulomb
interaction is $U_{\ell}=4.6~meV$. A barrier with width $7~nm$ leads
to $t_c=0.029~meV$, $U_{\ell,j}=2.2~meV$ and
$\eta=d_1/D=37/118=0.29$. Using these realistic physical parameters,
we obtain the heat current and heat rectification efficiency versus
temperature for $In_{0.6}Ga_{0.4}As/GaAs$ QD junction and show them
as triangle marks in Fig. 5. Compared to the dotted line, the heat
current is suppressed, whereas the rectification efficiency is
enhanced. This result indicates that the heat current  is not a
monotonic increasing function of $U_{\ell,j}$. Other systems of
interest include Ge$_x$Si$_{1-x}$ QDs embedded in Si nanowires in
which the confined carriers are holes.

In the presence of phonon heat current, the rectification efficiency
obtained above will be suppressed. To estimate the phonon heat
current, we calculate the thermal conductivities of nanowires  by
using the method and physical parameters given in our previous work
for studying semiconductor quantum wells.$^{21}$ The low temperature
phonon thermal conductivity of nanowire was reported in Ref. 22 for
studying the thermoelectric properties of Kondo insulator nanowires.
Figure 6(a) shows the calculated phonon thermal conductivity
($\kappa_{ph}$) of silicon and $GaAs$ nanowires with lateral
dimension $a=30~nm$ as a function of temperature. In diagram (b)
$\kappa_{ph}$ of silicon nanowire is smaller than that of GaAs
nanowire in the low temperature regime. Fig. 6(c) shows the quantum
size effect on $\kappa_{ph}$ of silicon and $GaAs$ nanowires at
$T=3K$. The results of Fig. 6(c) imply that the electron heat
rectification behavior as shown in Fig. 5 is easier to observe at
low temperatures in silicon system than in $GaAs$ system. We obtain
a phonon heat current $Q_{ph}=1.2\times 10^{-11}W$ for a single
silicon nanowire with cross-sectional area $A=(30~nm)^2$ and length
$L=150~nm$ at $3K$. This indicates that the results of Fig. 5 will
be washed out in the presence of phonon heat current. To reduce the
phonon heat current, we design a special structure as shown in the
inset of Fig.~6(a) with a smaller cross-sectional area $A=(5~nm)^2$,
which can be realized by advanced nanofabrication technique. The
phonon heat current of silicon nanowire can reduce to about
$Q_{ph}=30~fW$ for $A=(5~nm)^2$, which is closes to the heat
current, $Q\approx 10 Q_0=38.6~fW$ considered in Fig. 5. It is
conceivable that the phonon thermal conductivity of a nanowire can
be further reduced by the interface scattering effect due to the
presence of QDs.$^{23}$ Therefore, $Q_{ph}$ of silicon nanowire in
the presence of Ge$_x$Si$_{1-x}$ QDs can be smaller than $10~Q_0$ at
$3 K$. So far, many theoretical mechanisms employing phonon or
photon carriers for thermal rectifiers have been proposed.$^{24-26}$
However, few literatures have reported thermal rectification effect
experimentally.$^{27}$

\begin{flushleft}
{\bf Summary}\\
\end{flushleft}
The heat current of SCQD system including DQD and TQD under finite temperature bias has been investigated theoretically.
It is shown the thermal rectification behavior can arise from the electron  energy transport in the absence of
phonon heat current. Compared to a single QD with two levels$^{19}$ or
parallel multiple QDs,$^{13}$ the condition for thermal
rectification behavior of SCQD system is easier to realize. For example, one could consider the fabrication method of Ref. 9. The
presence of phonon heat current will seriously suppress the thermal rectification efficiency. Thus, reducing the phonon
heat current via smart nanostructure design becomes a key issue to the observation of direction-dependent
electron heat current.

\begin{flushleft}
{\bf Acknowledgments}\\
\end{flushleft}
This work was supported in part by the National Science Council of
the Republic of China under Contract Nos. NSC
101-2112-M-008-014-MY2, and NSC 101-2112-M-001-024-MY3.

\mbox{}\\
E-mail address: mtkuo@ee.ncu.edu.tw\\
E-mail address: yiachang@gate.sinica.edu.tw\\

\mbox{}

\newpage

\section*{Figure captions}
\begin{figure}[h]
\includegraphics[angle=0, scale=0.3]{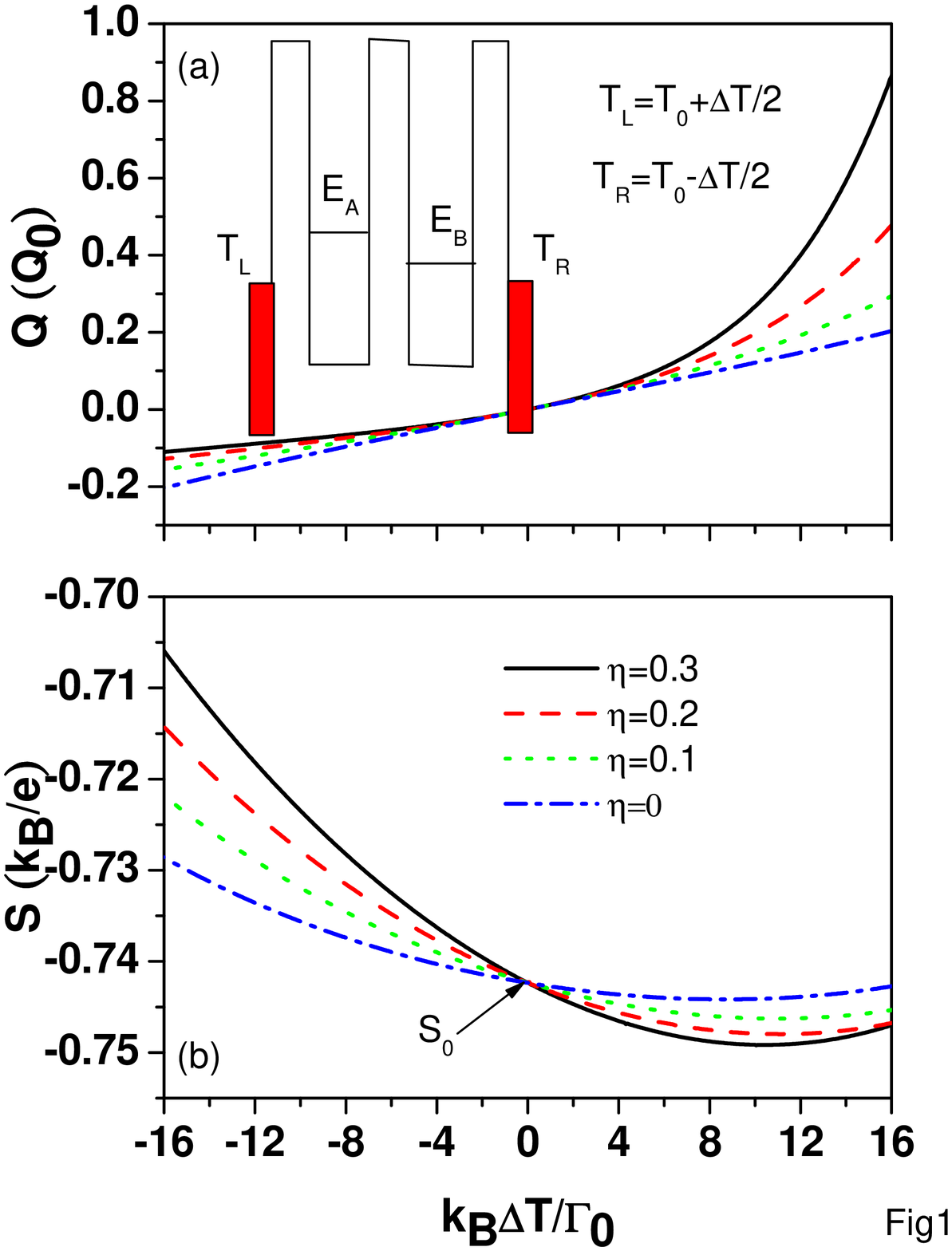}
\centering \caption{(a) Heat current, and (b) Seebeck coefficient as
a function of temperature difference $\Delta T$ for different $\eta$
factors. $E_A=E_F+20\Gamma_0$, $E_B=E_F+10\Gamma_0$.
$U_{\ell}=300\Gamma_0$, $U_{\ell,j}=20\Gamma_0$,
$t_{\ell,j}=1\Gamma_0$, and $\Gamma_L=\Gamma_R=3\Gamma_0$. Note that
$Q_0=\Gamma^2_0/h$.}
\end{figure}

\begin{figure}[h]
\includegraphics[angle=0, scale=0.3]{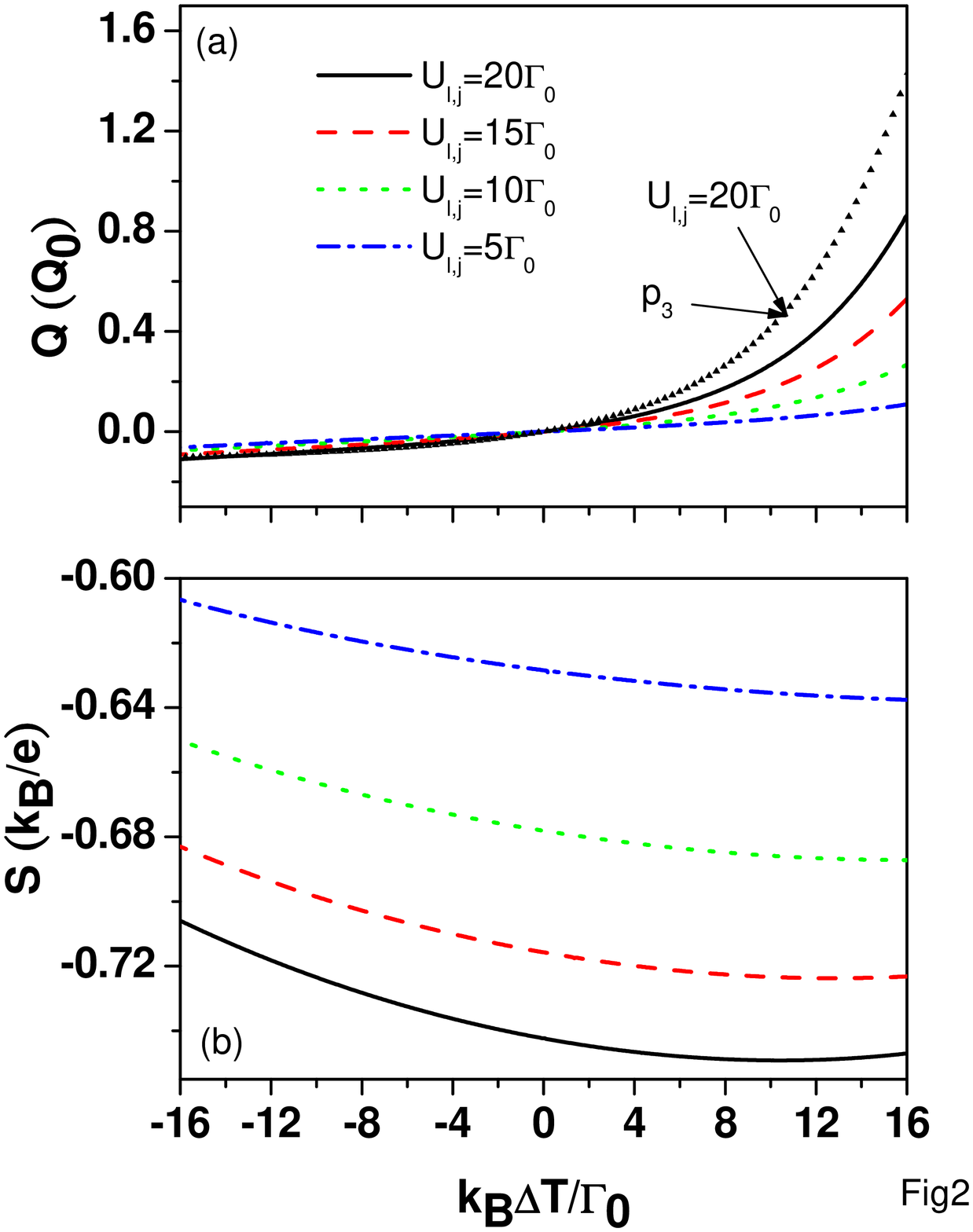}
\centering \caption{(a) Heat current and (b) Seebeck coefficient as
a function of temperature difference $\Delta T$ for different
interdot Coulomb interactions. Other physical parameters are the
same as those in Fig.~1(a) for $\eta=0.3$. The curve labeled by
$p_3$ considers only the $p_3$ configuration in calculating the heat
current.}
\end{figure}
\begin{figure}[h]
\includegraphics[angle=0, scale=0.3]{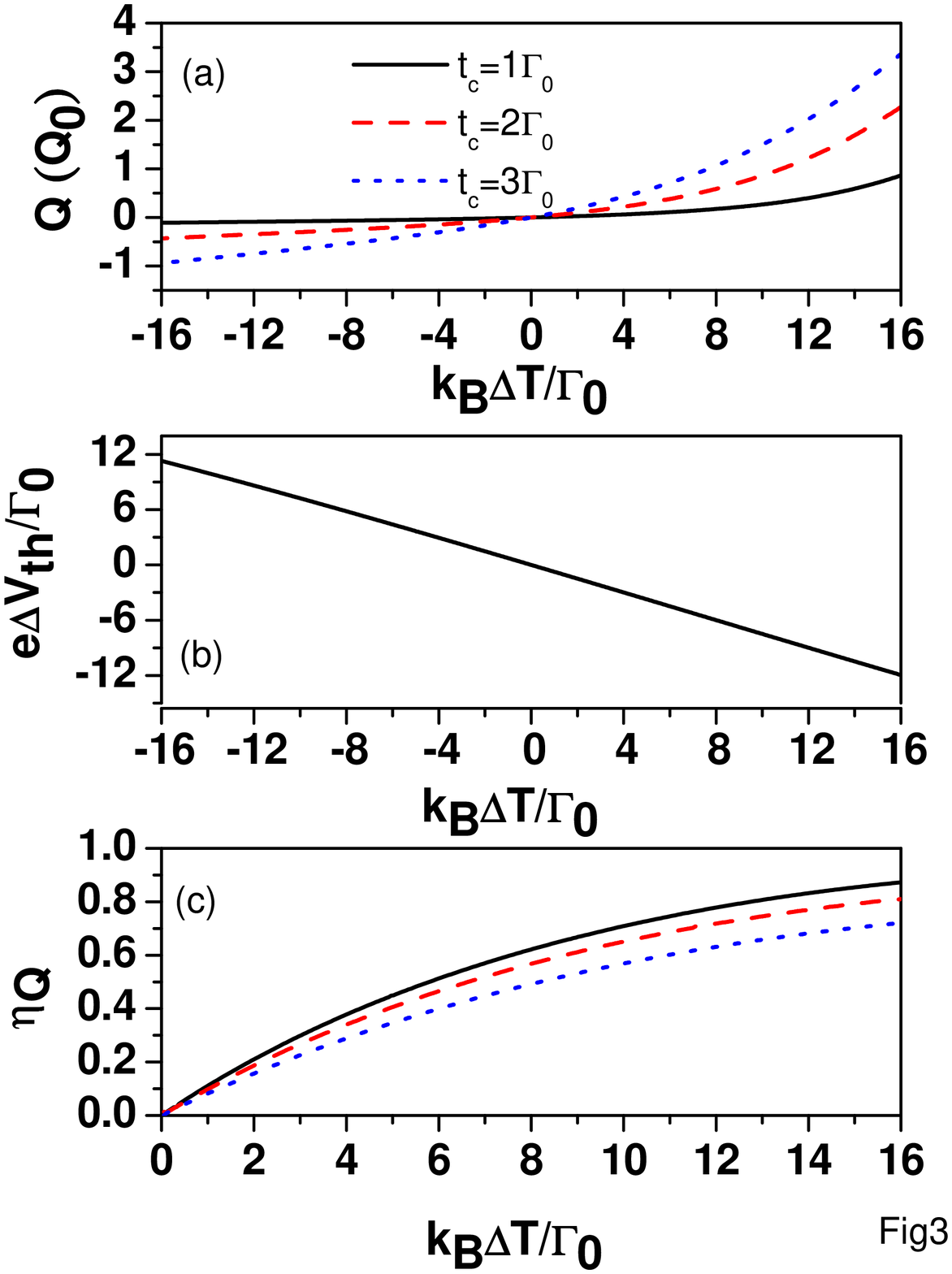}
\centering \caption{(a) Heat current (Q), (b) thermal voltage
($\Delta V_{th}$) and (c) thermal rectification efficiency
($\eta_Q$) as a function of temperature difference $\Delta T$ for
different interdot hopping strengths. Other physical parameters are
the same as those in Fig. 1(a) for $\eta=0.3$.}
\end{figure}
\begin{figure}[h]
\includegraphics[angle=0, scale=0.3]{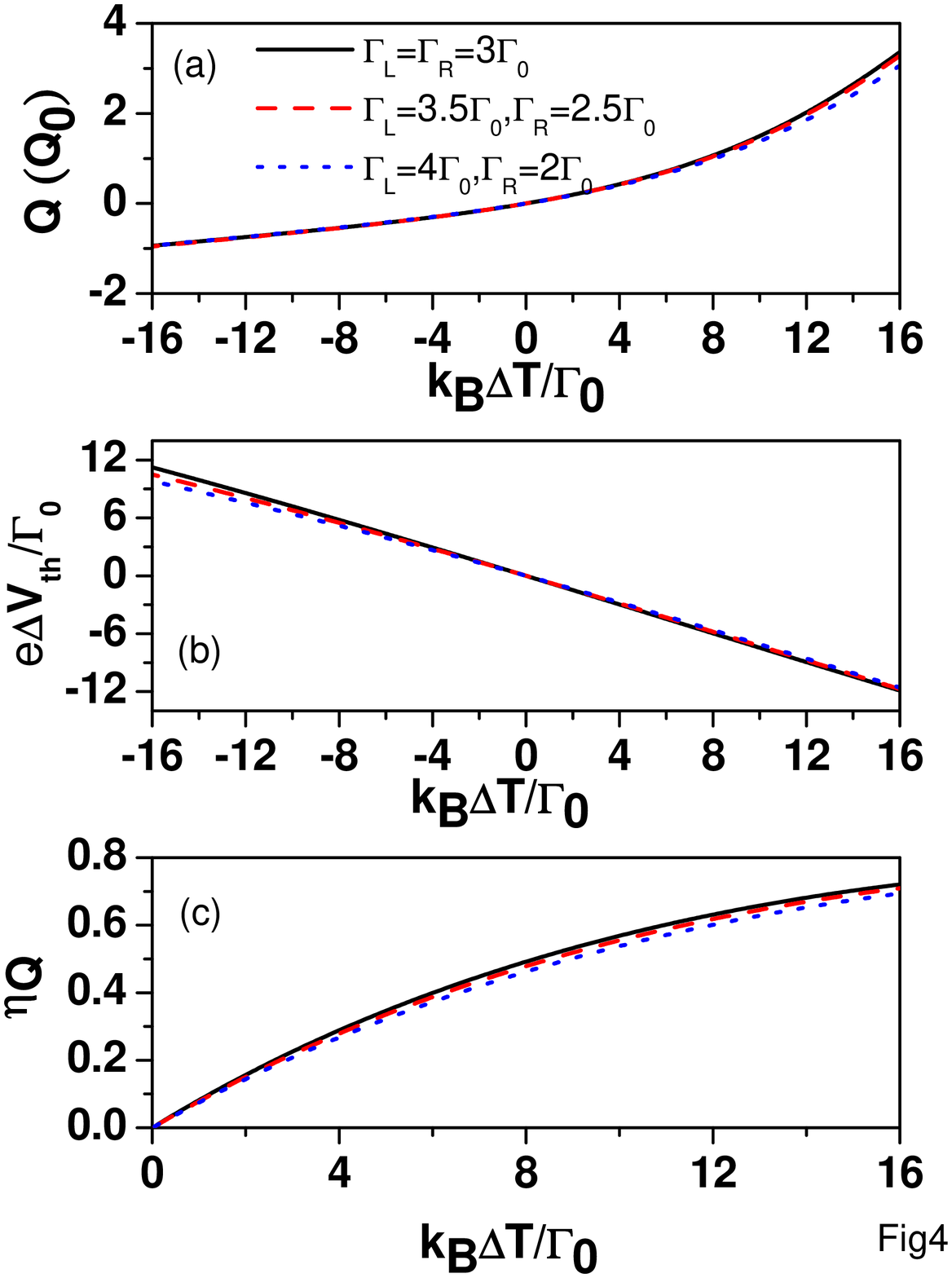}
\centering \caption{(a) Heat current (Q), (b) thermal voltage
($\Delta V_{th}$) and (c) thermal rectification efficiency
($\eta_Q$) as a function of temperature difference $\Delta T$ for
different tunneling rates at $\Gamma_L+\Gamma_R=3\Gamma_0$ and
$t_c=3\Gamma_0$. Other physical parameters are the same as those in
Fig. 1(a) for $\eta=0.3$.}
\end{figure}
\begin{figure}[h]
\includegraphics[angle=-90, scale=0.3]{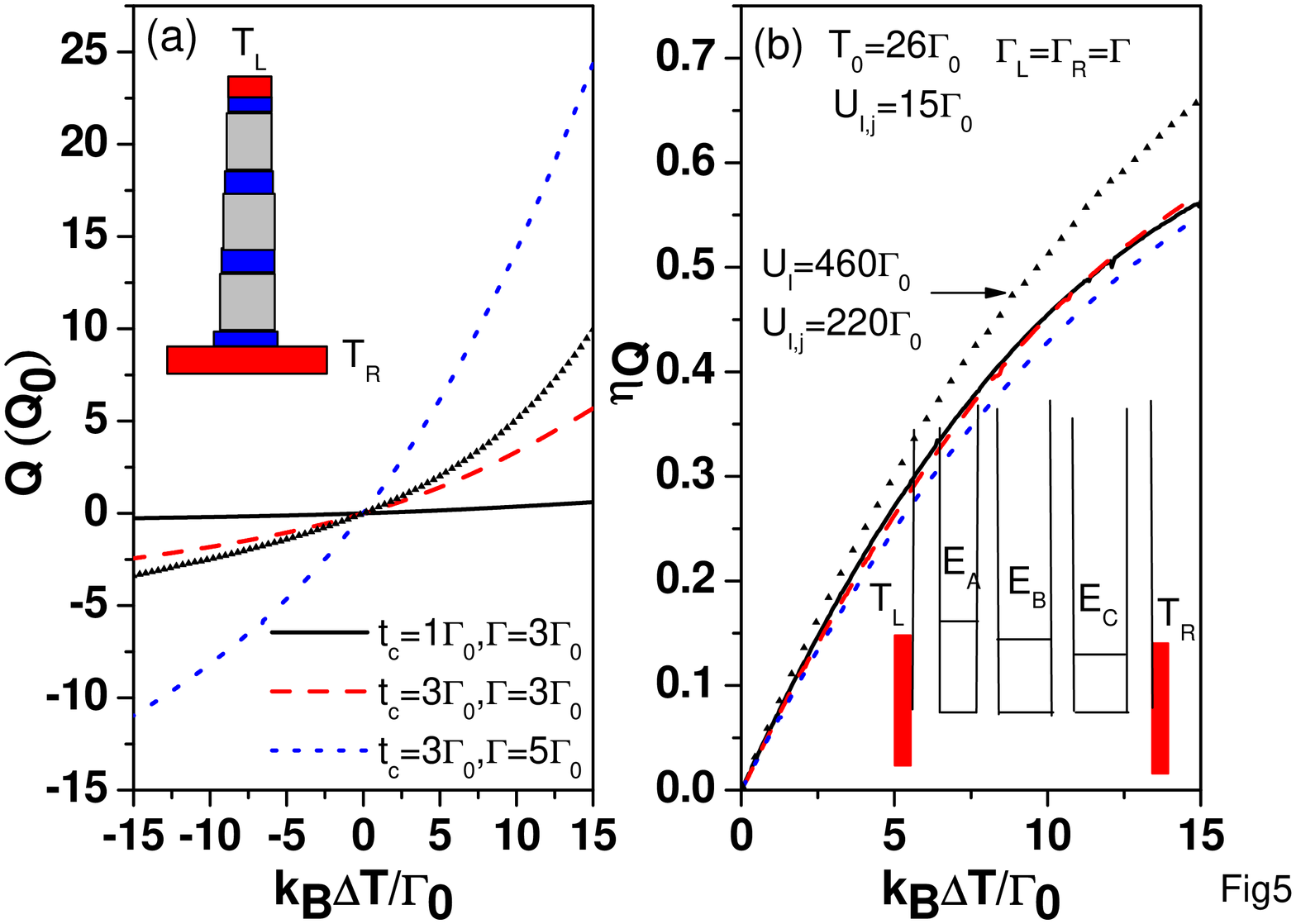}
\centering \caption{(a) Heat current (Q),  and (b)  thermal
rectification efficiency ($\eta_Q$) as a function of temperature
difference $\Delta T$ for different physical parameters in the case
of TQDs with $U_{\ell}=300\Gamma_0$, $U_{\ell,j}=15\Gamma_0$,
$t_{\ell,j}=t_c$, and $\Gamma_L=\Gamma_R=\Gamma$. If
$\Gamma_0=10~\mu eV$, $Q_0=\Gamma^2_0/h=3.86~fW$. The curve with
triangle marks is calculated by using $U_{\ell}=4.6~meV$,
$U_{\ell,j}=2.2~meV$ and $\Gamma=0.07~meV$.}
\end{figure}
\begin{figure}[h]
\includegraphics[angle=-90, scale=0.3]{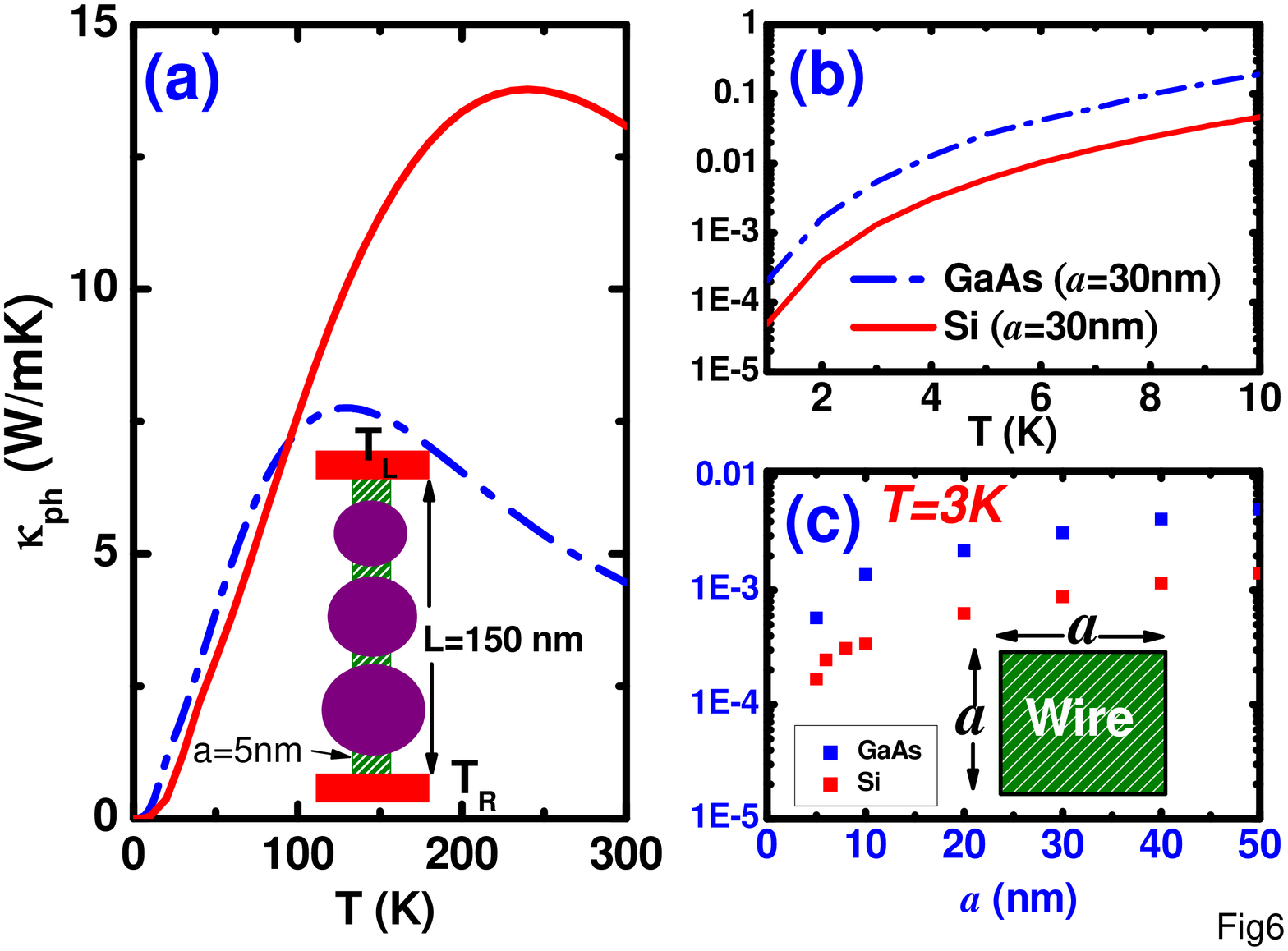}
\centering \caption{The thermal conductivities ($\kappa_{ph}$) as a
function of temperature for silicon and $GaAs$ nanowires with cross
section $A=(30~nm)^2$. Diagrams (a) and (b) show, respectively, the
$\kappa_{ph}$ in the high and low temperature regimes.  Fig. 6(c)
shows the phonon thermal conductivity as a function of diameter for
rectangular silicon and $GaAs$ nanowires at 3 $K$.}
\end{figure}


\begin{thebibliography}{50}


\bibitem[1]{Guo}L. J. Guo, E. Leobandung and S. Y. Chou, Science. \textbf{275}, 649
(1997).

\bibitem[2]{Ish} H. Ishikuro and T. Hiramoto, Appl. Phys. Lett. \textbf{71}, 3691(1997).

\bibitem[3]{Joa} C. Joachim, J. K. Gimzewski and A. Aviram, Nature
\textbf{408},
541 (2000).

\bibitem[4]{kuo1} D. M. T. Kuo, and Y. C. Chang,
Phys. Rev. Lett. \textbf{99}, 086803 (2007).

\bibitem[5]{Cha} Y. C. Chang and D. M. T. Kuo, Phys Rev B \textbf{77}, 245412 (2008).

\bibitem[6]{Han} R. Hanson, L. P. Kouwenhoven, J. R. Petta, S. Tarucha, and L. M. K. Vandersypen, Rev. Mod. Phys. \textbf{79},
1217 (2007).

\bibitem[7]{Ono} Ono K, Austing DG, Tokura Y, Tarucha S
\textit{Science}  \textbf{297}, 1313 (2002).

\bibitem[8]{Bus} M. Busl, G. Granger, L. Gradreau, R. Sanchez, A. Kam,
M. Pioro-Ladriers, S. A. Studenikin, P. Zawadzki, Z. R. Wasilewski,
A. S. Sachrajda and G. Platero, Nature Nanotechnology \textbf{7}, 1
(2013).

\bibitem[9]{Ama} S. Amaha, W. Izumida, S. Teraoka, S. Tarucha, J.
A. Gupta and D. G. Austing, Phys. Rev. Lett. \textbf{110}, 016803
(2013).

\bibitem[10]{Bra} F. R. Braakman, P. Barthelemy, C. Reichi, W.
Wegscheider, and L. M. K. Vandersypen, Appl. Phys. Lett. \textbf{
102}, 112110 (2013).

\bibitem[11]{Zeb} M. Zebarjadi, K. Esfarjania, M.S. Dresselhaus, Z.F. Ren and G.
Chen, Energy Environ Sci \textbf{5}, 5147 (2012).


\bibitem[12]{Mur} P. Murphy, S. Mukerjee, and J. Moore,
 Phys. Rev. \textbf{B} \textbf{78}, 161406 (2008).

\bibitem[13]{Kuo2} D. M. T. Kuo and  Y. C. Chang,
 Phys. Rev. B \textbf{81}, 205321 (2010).

\bibitem[14]{Liu} J. Liu, Q. F. Sun, and X. C. Xie,
Phys. Rev. B \textbf{81}, 245323 (2010).

\bibitem[15]{Kuo3} D. M. T. Kuo, S. Y. Shiau and Y. C. Chang,
Phys. Rev. B \textbf{84}, 245303 (2011).

\bibitem[16]{Kuo4} D. M. T. Kuo and Y. C. Chang,
Nanoscale Res. Lett. \textbf{7}, 257 (2012).

\bibitem[17]{Kuo5} D. M. T. Kuo and Y. C. Chang, Nanotechnology \textbf{24}, 175403 (2013) and
arXiv:1209.0506.v3


\bibitem[18]{Hau} H. Haug and A. P. Jauho, Quantum Kinetics in Transport and Optics
of Semiconductors (Springer, Heidelberg, 1996).

\bibitem[19]{Sch} R. Scheibner, M. Konig, D. Reuter, A. D.Wieck, C. Gould, H.
Buhmann and L. W. Molenkamp, New. J. Phys. \textbf{10}, 083016
(2008).

\bibitem[20]{Kuo6} D. M. T. Kuo and Y. C. Chang, Phys. Rev. B
\textbf{61}, 11051 (2000).

\bibitem[21]{Tse} Y. C. Tseng, D. M. T. Kuo and Y. C. Chang, J.
Appl. Phys. \textbf{113}, 113706 (2013).

\bibitem[22]{Zha} Y. Zhang, M. S. Dresselhaus, Y. Shi, Z. Ren, and
G. Chen, Nano Lett. \textbf{11}, 1166 (2011).

\bibitem[23]{Nik} D. L. Nika, E. P. Pokatilov, A. A. Balandin, V. M. Formin, A.
Rastelli, and O. G. Schmidt, Phys. Rev. B\textbf{ 84}, 165415
(2011).

\bibitem[24]{Li} B. Li, L. Wang and G. Casati, Phys. Rev. Lett. \textbf{93} 184301 (2004).

\bibitem[25]{Hu} B. B. Hu, L. Yang, and Y. Zhang
Phys. Rev. Lett.\textbf{ 97} 124302  (2006).

\bibitem[26 ]{Ote} C. R. Otey, W. T. Lau, and S. H. Fan, Phys. Rev. Lett. \textbf{104} 154301
(2010).

\bibitem[27]{Cha1} C. W. Chang, D. Okawa1, A. Majumdar, and A. Zettl, Science \textbf{17}
1121 (2006).






\end{thebibliography}
\end{document}